\documentclass[%
 aip,
 amsmath,amssymb,
 reprint,%
]{revtex4-1}
\usepackage{chemformula} 
\usepackage{bm}
\usepackage{amsmath}
\usepackage{amssymb}
\usepackage{amsthm}
\usepackage{amsfonts}
\usepackage{graphicx}
\usepackage{makeidx}
\usepackage{color}
\usepackage{chemformula} 
\usepackage[T1]{fontenc} 

\usepackage[utf8]{inputenc}
\usepackage[T1]{fontenc}
\usepackage{mathptmx}

\begin{document} 

\title{Epitaxial stabilization of (111)-oriented frustrated quantum pyrochlore thin films}

\author{Fangdi Wen}
\email{fangdi.wen@rutgers.edu}
\affiliation{Department of Physics and Astronomy, Rutgers University, Piscataway, New Jersey 08854, USA}

\author{Tsung-Chi Wu}
\affiliation{Department of Physics and Astronomy, Rutgers University, Piscataway, New Jersey 08854, USA}

\author{Xiaoran Liu}
\affiliation{Department of Physics and Astronomy, Rutgers University, Piscataway, New Jersey 08854, USA}

\author{Michael Terilli}
\affiliation{Department of Physics and Astronomy, Rutgers University, Piscataway, New Jersey 08854, USA}

\author{Mikhail Kareev}
\affiliation{Department of Physics and Astronomy, Rutgers University, Piscataway, New Jersey 08854, USA}

\author{Jak Chakhalian}
\affiliation{Department of Physics and Astronomy, Rutgers University, Piscataway, New Jersey 08854, USA}

\date{\today}

\begin{abstract}
 Frustrated rare-earth pyrochlore titanates, Yb$_2$Ti$_2$O$_7$ and Tb$_2$Ti$_2$O$_7$, have been proposed as promising candidates to realize quantum spin ice (QSI). 
 Multiple exotic quantum phases, including Coulombic ferromagnet, quantum valence-bond solid, and quadrupolar ordering, have been predicted to emerge in the QSI state upon the application of a (111)-oriented external magnetic field. 
 Here, we report on the primal successful layer-by-layer growth of thin films of frustrated quantum pyrochlores, R$_2$Ti$_2$O$_7$ (R = Er, Yb, and Tb), along the (111) direction. 
 We confirm their high crystallinity and proper chemical composition by a combination of methods, including in-situ RHEED, x-ray diffraction, reciprocal space mapping, and x-ray photoelectron spectroscopy. 
 The availability of large area (111)-oriented QSI structures with planar geometry offers a new complementary to the bulk platform to explore strain and magnetic field dependent properties in the quasi-2D limit. 
 
\end{abstract}
\maketitle
\section{INTRODUCTION}
In the context of modern condensed matter, there has been a growing interest in searching for exotic states and  excitations in highly frustrated materials systems. 
For example, the excitations of magnetic monopoles due to the fractionalized dipolar degrees of freedom can be found in classical spin ice (CSI) materials such as Ho$_2$Ti$_2$O$_7$ and Dy$_2$Ti$_2$O$_7$, the rare-earth pyrochlore oxides.\cite{fennell2009magnetic,castelnovo2008magnetic,jaubert2009signature,bramwell2009measurement,harris1997geometrical,den2000dipolar,bramwell2001spin,gardner2010magnetic,rau2019frustrated,castelnovo2012spin} 
Another intriguing state of correlated matter is quantum spin liquid (QSL), characterized by emergent fractionalized excitations with long-range entanglement. \cite{balents2010spin,broholm2020quantum,savary2016quantum,zhou2017quantum,knolle2019field,han2012fractionalized,janvsa2018observation} 

Recently, a new class of QSLs named quantum spin ice (QSI) has been theoretically proposed by including quantum fluctuations, which act to prevent the spins from reaching a symmetry-breaking ordered phase of CSIs. \cite{gingras2014quantum,rau2019frustrated,hermele2004pyrochlore,ross2011quantum,lee2012generic,benton2012seeing,ross2011quantum,savary2012coulombic,RevModPhys.88.041002} 
QSI has drawn much attention since it supports a QSL ground state described in the framework of compact lattice gauge theory with exotic excitations, including magnetic monopoles, U(1) gauge photons, and spinons. \cite{chen2017dirac,savary2012coulombic,hermele2004pyrochlore,RN424,PhysRevResearch.2.013334,PhysRevLett.80.2933,PhysRevLett.115.097202,PhysRevLett.113.117201} 
Interestingly, even if a QSI phase is not ultimately realized, due to its highly frustrated nature, proximity to the QSI state is anticipated to induce several unusual quantum phases. \cite{savary2012coulombic,lee2012generic,gingras2014quantum,wan2016spinon,kato2015numerical,benton2018quantum,udagawa2019spectrum,sibille2020quantum,RN424,Poole_2007}

Experimentally, many frustrated quantum pyrochlores, including R$_2$Ti$_2$O$_7$ (R = Tb and Yb), have been put forward as promising candidates for QSI. \cite{rau2019frustrated,pan2014low,ross2011quantum,molavian2007dynamically,ruff2007structural,tokiwa2018discovery,sibille2018experimental,benton2018instabilities,PhysRevB.94.205107,PhysRevLett.115.097202,PhysRevLett.113.117201,Scheie202008791,Takatsu_2011,Molavian_2009} 
With the localized nature of \textit{f}-shell moments as well as strong spin-orbit interaction, these \textit{f}-electron systems exhibit highly directional exchange interactions akin to those found in Kitaev magnets, resulting in strongly frustrated behavior. \cite{rau2019frustrated,nasu2014vaporization,hickey2019emergence,takagi2019concept,motome2020hunting} 
From the experimental viewpoint, however, more often than not, the exact nature of the ground state and low-energy excitations in these quantum pyrochlores remains largely unresolved. 
One reason is the lack of detailed knowledge on the role of disorder. \cite{savary2017disorder,martin2017disorder,wen2017disordered,benton2018instabilities,bowman2019role} 
To illustrate, recently it was found that disorder can promote a new phase in the candidate QSI pyrochlore  Pr$_2$Zr$_2$O$_7$, which is disordered and yet exhibits short-range antiferro-quadrupolar correlations and mimics the QSL-like features in neutron scattering.  \cite{martin2017disorder,petit2016antiferroquadrupolar} 
Another challenge is the difficulty of fabricating high-quality single-crystalline samples with low disorders, which hinders the intrinsic QSI physics and unavoidably gives rise to inconsistency between experiments. \cite{van1990powder,kumar2007high,scott2011high,kermarrec2015gapped,vlavskova2020high,ruminy2016sample,ferromagnetic2003YY,Yaouanc_2016} 
To address these challenges, it is thus critical to develop new methods of materials synthesis that are distinct from the conventional solid-state synthesis route to unveil the true nature of frustrated quantum pyrochlores.

One compelling approach that has attracted significant recent  interest is to grow the pyrochlore materials as thin-films oriented along the (111) direction.  \cite{liu2016geometrical,chakhalian2020strongly,liu2019emergent,liu2020situ}
The resulting films contain alternating kagome and triangular atomic planes of magnetically active rare-earth ions, which are naturally formed in such orientation and are known to support emergent magnetic states, including QSL.  \cite{liu2016geometrical,chakhalian2020strongly,liu2019emergent,liu2019quantum} 
Based on this motif, many exotic phenomena including quantum kagome valence bond solid, Coulombic ferromagnet, quadrupolar, and monopole super-solids  were proposed to emerge in the QSI films under a magnetic field applied along the (111) direction. \cite{bojesen2017quantum,kadowaki2018dimensional} 
Surprisingly, to date, the synthesis of (111) pyrochlore thin films remains very limited and often relies on access to the commercially unavailable pyrochlore substrate Y$_2$Ti$_2$O$_7$. \cite{bovo2016layer,bovo2014restoration} 
Besides, the epitaxial control of the QSI films can be vital for potential applications in the subfield of quantum information technology.

Here, we report on the primal successful layer-by-layer growth of high-quality thin films of a series of frustrated quantum pyrochlores, R$_2$Ti$_2$O$_7$ (R = Er, Tb and Yb), on the 5$\times$5 mm$^2$ (111) yttria-stabilized ZrO$_2$ (YSZ) substrate. 
We confirm the high crystallinity of thin films by in-situ reflected high  energy  electron diffraction (RHEED), x-ray measurements including x-ray diffraction (XRD), and reciprocal space map (RSM), and validate that all the films exhibit correct chemical composition by x-ray photoelectron spectroscopy (XPS). 
Our work offers a complementary to the bulk route to resolve the puzzles in physics of frustrated quantum pyrochlores and contributes to the possible realization of the long-thought QSI state.

\begin{figure*}[t]
\vspace{-0pt}
\includegraphics[width=\textwidth]{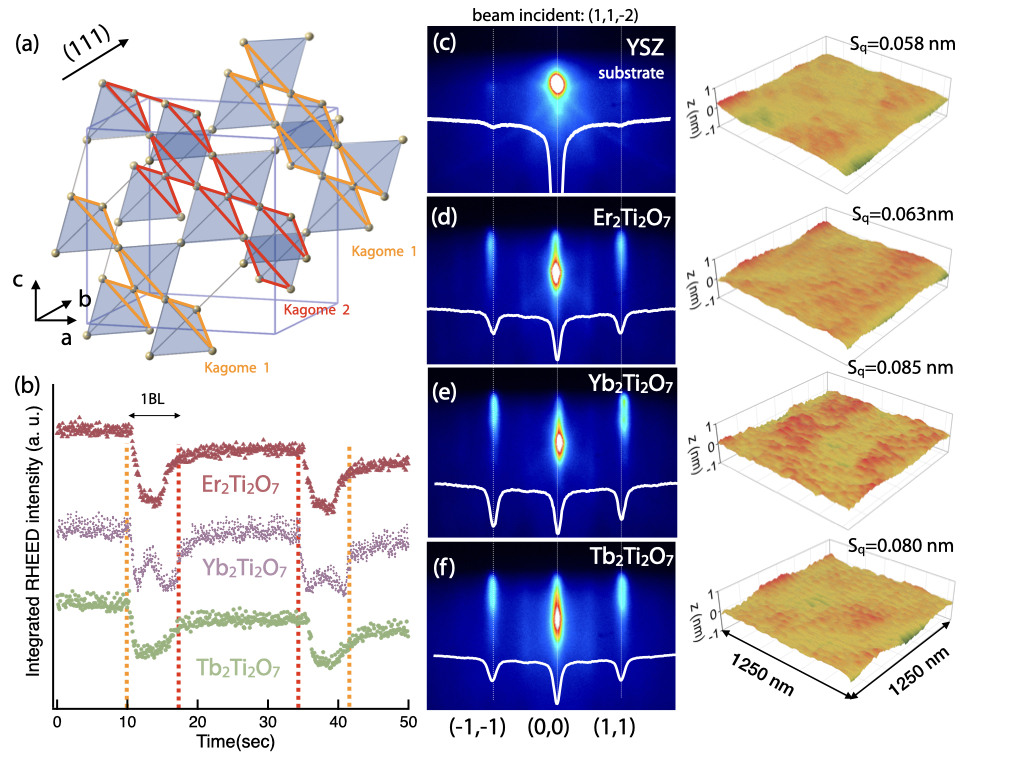}
\caption{\textbf{(a)} Schematic picture of R site sublattice in R$_2$Ti$_2$O$_7$. \textbf{(b)} RHEED oscillations during the high-frequency growth, where each interval exhibits the growth of one BL. Note, only the kagome layers are marked for clarity. \textbf{(c)-(f)}   RHEED picture of YSZ, Er$_2$Ti$_2$O$_7$, Yb$_2$Ti$_2$O$_7$, and Tb$_2$Ti$_2$O$_7$, respectively. Electron beam incident from the (1,1,-2) crystal orientation of YSZ. The white inset curve is a horizontal intensity line cut of the corresponding figure. To the right of (c), (d), (e), and (f) are the AFM of YSZ, Er$_2$Ti$_2$O$_7$, Yb$_2$Ti$_2$O$_7$, and Tb$_2$Ti$_2$O$_7$ surface. With the same scanning area of 1.25$\mu$m $\times$ 1.25$\mu$m, the corresponding root mean square roughness obtained by AFM is: YSZ - 58pm, Er$_2$Ti$_2$O$_7$ - 63pm, Yb$_2$Ti$_2$O$_7$ - 85pm, and Tb$_2$Ti$_2$O$_7$ - 80pm.}
\label{RHEED}
\end{figure*}

\section{RESULTS and DISCUSSION}

Three frustrated quantum pyrochlores, R$_2$Ti$_2$O$_7$ (R =Er, Tb, and Yb), were epitaxially stabilized on YSZ substrates using pulsed-laser deposition with in-situ RHEED control at the identical growth condition. 
Specifically, before the growth, substrates were heated up to 750 $^\circ$C at 10 $^\circ$C/min under 120 mTorr oxygen pressure. 
During the growth, the presence of the layer-by-layer growth mode was determined by high-pressure RHEED. 
All the reported films were deposited at the 18 Hz repetition rate interrupted by 15-second intervals. 
A high-frequency deposition enlarges the supersaturation limit, lowers the nucleation barrier, and thus markedly improves the growth. \cite{kareev2011sub} 
After the growth, samples were kept at the growth condition for 10 mins, and then cooled down to room temperature with the ramp rate of 15 $^\circ$C/min. It is noteworthy that no annealing was required in this process.

\begin{figure*}[t]
\vspace{-0pt}
\includegraphics[width=\textwidth]{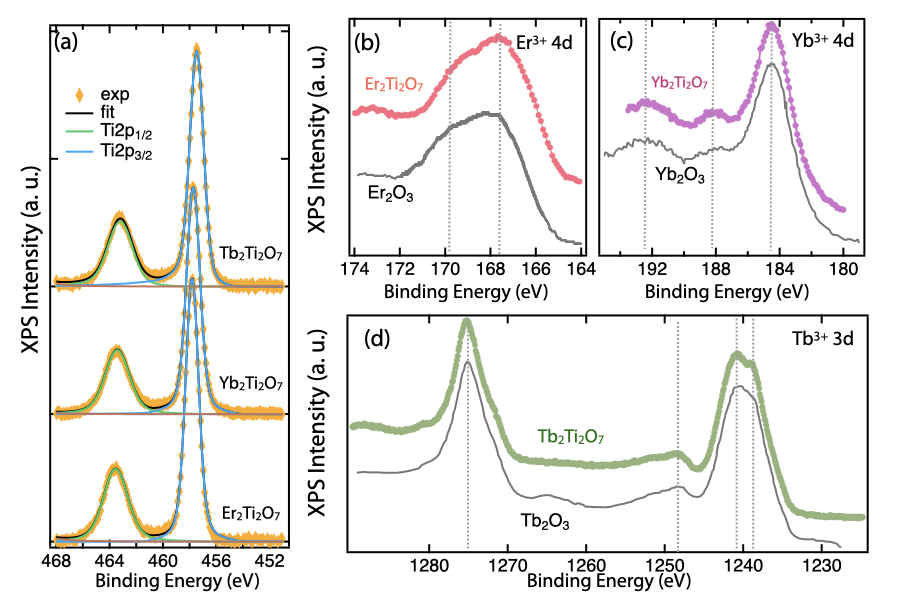}
\caption{ \textbf{(a)} XPS of Ti 2p on Er$_2$Ti$_2$O$_7$, Yb$_2$Ti$_2$O$_7$ and Tb$_2$Ti$_2$O$_7$ films. All the background has been subtracted for better comparison. \textbf{(b)} XPS of Er 4d in Er$_2$Ti$_2$O$_7$ sample in comparison to the data adapted from the reported Er$_2$O$_3$.\cite{Er2O3_XPS_4} \textbf{(c)} XPS of Yb 4d in Yb$_2$Ti$_2$O$_7$ sample in comparison to the data adapted from the reported Yb$_2$O$_3$.\cite{YTO_XPS_3} \textbf{(d)} XPS of Tb 3d in Tb$_2$Ti$_2$O$_7$ sample in comparison to the data adapted from the reported Tb$_2$O$_3$.\cite{TTO_XPS_1}}
\label{XPS}
\end{figure*}

Surprisingly,  as shown in Fig. \ref{RHEED}(b), unlike a conventional  RHEED pattern expected for interrupted growth, \textit{two oscillations} were found to appear for all samples albeit with a different number of pulses required to complete a cycle. 
This unexpected result can be readily explained by the fact that since the rare-earth sites and Ti sites have the same sublattice structure (but with a small displacement), a kagome layer is composed of rare-earth ions which are coplanar with a triangle layer of Ti$^{4+}$ ions. 
Therefore, a complete structural unit requires two inequivalent bilayers (BL) composed of one rare-earth kagome plane (K$_R$) plus one triangle Ti ions plane  (T$_T$) or BL = K$_R$ - T$_T$  and another bilayer composed  of kagome Ti (K$_T$) and triangle rare-earth (T$_R$) planes, thus overall 1 u.c.= K$_R$-T$_T$-K$_T$-T$_R.$ 
As illustrated in Fig. 1(a), these BLs are not precisely on the top of each other but rather stacked with a small offset . 
Further thickness analysis revealed that between each interval, we indeed deposited two BLs of material.
Furthermore, this structural consideration is in good agreement with the observed doublet structure of RHEED intensity in all (111) oriented films. 
The ability to see the two-peak structure in RHEED oscillations during high-frequency deposition provides strong evidence for the layer-by-layer growth. 

After each deposition, RHEED intensity recovers to almost the same intensity, with a minor decay corresponding to the natural decay of the LaB$_6$ electron source. 
This result implies that the surface roughness remained roughly constant for each unit  cell.
The RHEED patterns of all samples, including the YSZ substrate, are shown in the left part of Fig. \ref{RHEED} (c)-(f). 
Because of the small incident angle, the (-1,-1) and (1,1) crystal peaks of YSZ were barely observed. 
However, soon after the initial deposition, two evident streaky peaks emerge at the (-1,-1) and (1,1) positions. 
A clear half-order reflections can be seen after the deposition, which is as expected since the pyrochlore lattice constant is twice of that of YSZ. 
When cooled down to 300 $^\circ$C, the RHEED pattern remained unchanged and disappeared at the room temperature due to the strong charging effect. 

\begin{figure*}[t]
\vspace{-0pt}
\includegraphics[width=\textwidth]{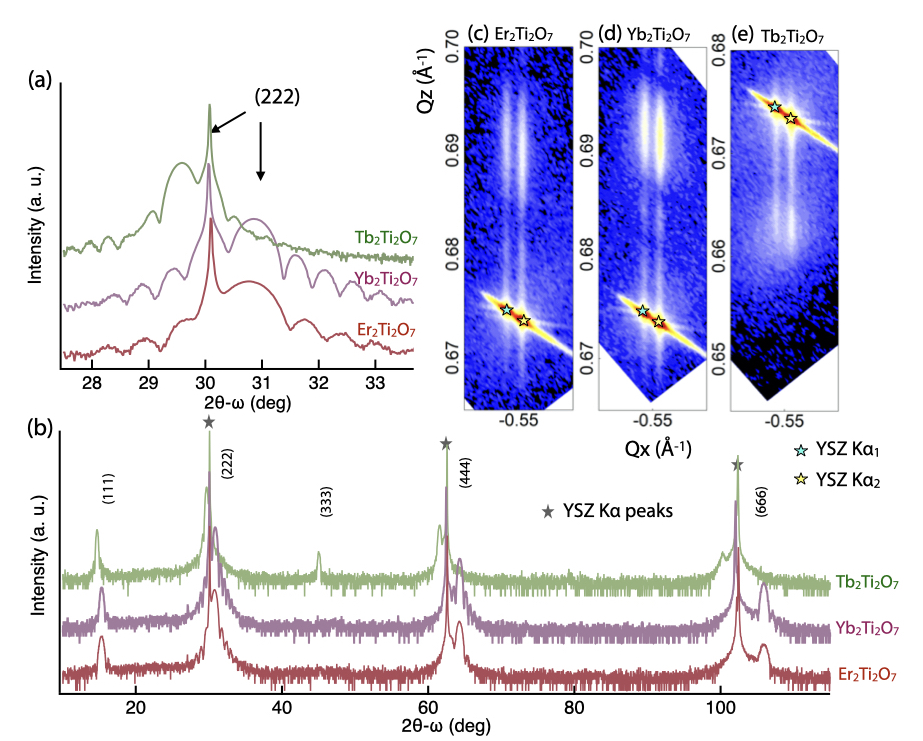}
\caption{\textbf{(a)} XRD of Er$_2$Ti$_2$O$_7$, Yb$_2$Ti$_2$O$_7$ and Tb$_2$Ti$_2$O$_7$ films near the film peak (222) and substrate peak (111). \textbf{(b)} Full range XRD of Er$_2$Ti$_2$O$_7$, Yb$_2$Ti$_2$O$_7$ and Tb$_2$Ti$_2$O$_7$. (c), (d), and (e) are the reciprocal space map of Er$_2$Ti$_2$O$_7$, Yb$_2$Ti$_2$O$_7$ and Tb$_2$Ti$_2$O$_7$ around the substrate peak (0,4,2) and film peak (0,8,4). Both K $\alpha_{1}$ and K $\alpha_{2}$ were detected, and we calculated Q vectors based on K $\alpha_{1}$ with wavelength of (1.5406\AA). All graphs show data acquired from three samples: Er$_2$Ti$_2$O$_7$ sample with thickness 16nm, Yb$_2$Ti$_2$O$_7$ sample with thickness 20nm, and Tb$_2$Ti$_2$O$_7$ sample with thickness 30nm.}
.
\label{XRD}
\end{figure*}

After the growth, each sample was characterized in four aspects: the surface morphology, crystallinity, chemical composition, and crystal structure. 
Other than determining the surface morphology by in-situ RHEED, atomic force microscopy (AFM) is another commonly-used probe. 
As seen in the right half of Fig. \ref{RHEED} (c), the YSZ substrate has a root mean square surface roughness S$_q$ of around 58 pm over a 1.25$\mu$m $\times$ 1.25 $\mu$m area.
The S$_q$ of Er$_2$Ti$_2$O$_7$ (d), Yb$_2$Ti$_2$O$_7$ (e) and Tb$_2$Ti$_2$O$_7$ (f) are all of the same order of 60-80 pm. 
Remarkably, since each BL corresponds to 80 pm, the surface roughness of all three films is within 1 BL, which is another manifestation of the atomically-flat surface due to the layer-by-layer growth.

To assure that during the deposition, ions are stabilized in the proper charge state, a detailed XPS analysis has been carried out on both  R (Er, Yb, and Tb)  and Ti core-states. 
During the XPS analysis, all the recorded data were calibrated to the carbon 1\textit{s} peak at 284.5 eV to eliminate the possible charging effect. 
The XPS core level data of Ti$^{4+}$ in Er$_2$Ti$_2$O$_7$, Yb$_2$Ti$_2$O$_7$ and Tb$_2$Ti$_2$O$_7$ are shown in Fig.\ref{XPS} (a). For better comparison the Shirley-type background was subtracted from all Ti XPS spectra. 
Ti 2\textit{p}$_{3/2}$ peak and Ti 2\textit{p}$_{1/2}$ peak were fitted with a fixed area ratio of 2:1. The measured binding energy for Ti 2\textit{p}$_{3/2}$ peak is around 458.8 eV, which is in good agreement with the reference Ti$^{4+}$ systems.\cite{shibagaki1999xps,wen2018evolution} 
Although titanium can easily lower its valency, the spectra show  no sign of Ti$^{3+}$ ($\sim$455 eV) present in any of the films.

On the other hand, being in the group of lanthanides, Er, Tb, and Yb almost exclusively form compounds with an oxidation state of +3.
Unlike the relatively simple lineshape of Ti$^{4+}$ XPS, the XPS spectra of R sites (Er, Yb, Tb) are significantly more complex. 
To overcome this issue, we compare our results to the reference oxide systems with well-defined charge states.
First, as shown in Fig 2(b), though the peak positions of Er$^{3+}$ vary in different materials systems, the XPS spectrum of Er$_2$O$_3$ shows a clear correspondence to our data. \cite{Er2O3_XPS_1,Er2O3_XPS_2,Er2O3_XPS_3} 
The observed fine structure of Er$^{3+}$ can be attributed to the hybridization of Eu ions into Er-O, Er-Er, and Er-Ti bonds and surface contamination. \cite{Er2O3_XPS_3} 
For Yb$_2$Ti$_2$O$_7$, the single 4\textit{d}$_{5/2}$ peak can be clearly seen at around 185 eV, which is in  good agreement with previous XPS result of Yb$^{3+}$ in Yb$_2$O$_3$ [see Fig. \ref{XPS} (c)]. \cite{wang2003manufacturing,pan2009physical,rahman2017fabrication} 
It is worthy to note that if there is Yb$^{2+}$ present in the material, an extra peak will show at around 181 eV, which is not the case for our result.
Finally, the spectrum of Tb$^{3+}$ 3\textit{d} spectrum is consistent with the XPS\ spectra reported for  Tb$_2$O$_3$.\cite{lee2020oxidation,cartas2014oxidation} At  this  point we can conclude that  in all  synthesised films show correct  R$^{3+}$  (i.e. Er$^{3+}$, Yb$^{3+}$, Tb$^{3+}$) and   Ti$^{4+}$ oxidation states. 

Next, we discuss the structural properties of the films. In the bulk, the reported lattice constants for Er$_2$Ti$_2$O$_7$ and Yb$_2$Ti$_2$O$_7$ are 10.075 \AA \ and 10.032 \AA, respectively.\cite{wang2019synthesis,arpino2017impact} 
Their lattice constants correspond to the spacing of 5.82 \AA \ for Er$_2$Ti$_2$O$_7$ and 5.79 \AA \ for Yb$_2$Ti$_2$O$_7$ for each BL grown along the (111) direction. 
However, from  the bulk result, it is known that the lattice constant of Tb$_2$Ti$_2$O$_7$ is very sensitive to the growth condition; specifically, depending on the synthesis technique, the lattice constant can vary from 10.127 \AA\ \cite{rule2009tetragonal} to above 10.15 \AA. \cite{gardner1998single,kumar2007high,van1990powder}
As clearly seen in Fig. \ref{XRD} (a), Er$_2$Ti$_2$O$_7$ and Yb$_2$Ti$_2$O$_7$ shows film peaks at higher angles than substrate peaks. 
The calculated spacing along the (111) direction is estimated to be 5.80 \AA \ for Er$_2$Ti$_2$O$_7$ and 5.78 \AA \ for Yb$_2$Ti$_2$O$_7$. 
The smaller out-of-plane lattice spacing indicates tensile strain in the film, which is expected as pyrochlores have the lattice constants smaller than that of the YSZ substrate (5.12\AA). 
Also, because Yb$_2$Ti$_2$O$_7$ has a more pronounced lattice mismatch with YSZ, it has a smaller out-of-plane spacing, which is confirmed by our XRD result. 

Notably, the (333) and (555) peaks are not observed in the Er$_2$Ti$_2$O$_7$ and Yb$_2$Ti$_2$O$_7$ films. 
If we define the peak ratio between (111) peak and (333) peak as \textit{r}, under cubic symmetry Fd$\bar{3}$m, $r_{\textrm{ETO}}$ =  99.56 and $r_{\textrm{YTO}}$ = 75.17, implying that the intensity of the (333) peak is indeed very low compared to the (111) peak. 
Moreover, even for the bulk  crystals of  Er$_2$Ti$_2$O$_7$ and Yb$_2$Ti$_2$O$_7$, the (333) peak is barely observed in neutron or x-ray scattering measurements. \cite{zhang2009synthesis,cao2014structure,baroudi2015symmetry} 
On the other hand, our XRD measurements on Tb$_2$Ti$_2$O$_7$ have revealed several variations in the structural properties, including the lattice constant and  intensity of the (333) peak. 
Here, assuming the Tb$_2$Ti$_2$O$_7$ is under Fd$\bar{3}$m space group, and $a_{\textrm{TTO}}$ = 10.15 \AA, the estimated $r_{\textrm{TTO}}$ = 96. 
However, our XRD result indicates that  the value of $r_{\textrm{TTO}}^{\text{exp}}$= 10.44, which is much smaller than expected for the Fd$\bar{3}$m space group.  
In addition, the estimated out-of-plane lattice spacing from (111), (222), (333) and (444) film peaks consistently yeilds \textit{d}$_{111}$= 6.03 \AA, which is by 2.6\% larger than $a_{\textrm{bulk}}$= 10.15 \AA  ~($d_{111}$=5.86 \AA). 
Based on the reciprocal space mapping (RSM) [see Fig. 3(c)], the film appear to be fully strained in-plane, therefore the experimental unit cell volume of Tb$_2$Ti$_2$O$_7$ is 1095 \AA$^3$, which is  4.7\% larger than that of the bulk unit cell (1045.7 \AA$^3$).
Despite the observed deviation in the lattice constant of  Tb$_2$Ti$_2$O$_7$, the XRD scans show  no extra peaks other than those of the (111) - oriented crystal, confirming the absence of a secondary chemical phase or domain separation in the Tb$_2$Ti$_2$O$_7$
films. 

As shown in Fig 3(a)-(b), all films demonstrate clear Kiessig fringes near the film peaks, which result from the interference of the x-ray beams reflected on the film surface and the interface between the film and the substrate. 
These fringes indicate that our films have high thickness homogeneity and small surface/interface roughness constant with the AFM data. 
With angle-dependent Kiessig fringes, the film thickness is estimated to be 16 nm for Er$_2$Ti$_2$O$_7$, 20 nm for Yb$_2$Ti$_2$O$_7$, and 30 nm for Tb$_2$Ti$_2$O$_7$. Furthermore, the RSM shown in Fig. \ref{XRD} (c) implies a perfect alignment between the substrate  (042) and film peaks (084) horizontally, signifying that all samples are fully-strained. 
The RSM data also corroborate that the out-of-plane lattice constants of Er$_2$Ti$_2$O$_7$ and Yb$_2$Ti$_2$O$_7$ are larger than the out-of-plane lattice constant of YSZ substrate, and smaller for  Tb$_2$Ti$_2$O$_7$.  
Altogether, we observed a unique crystal structure in Tb$_2$Ti$_2$O$_7$ film that deviates from the commonly-observed in the bulk Fd$\bar{3}$m cubic structure.
The origin of the enlarged unit-cell volume and how the change in the crystal structure may affect its physical properties remain an open question to be studied in the future.

\section{Conclusion}

In conclusion, we have epitaxially grown new high-quality frustrated quantum pyrochlores R$_2$Ti$_2$O$_7$ (R = Er, Tb, and Yb)  on (111)-oriented YSZ substrates in the layer-by-layer mode. 
All systems exhibit correct chemical composition on both rare-earth sites and Ti site. Structurally, all films are fully strained with high homogeneity and very small surface roughness. 
Among the three systems, Er$_2$Ti$_2$O$_7$ and Yb$_2$Ti$_2$O$_7$ demonstrate tensile lattice strain as expected in the assumption of tetragonal  distortion, while Tb$_2$Ti$_2$O$_7$ shows an unexpectedly larger lattice parameter for the Fd$\bar{3}$m space group. 
Our work not only lays a solid ground for highly frustrated lattices grown from pyrochlore materials into thin-film forms in a (111) direction, but also offers a novel materials platform to explore the effects of strain and magnetic field in the  quasi-2D limit with the unique potential for the possible realization of QSI.

\section{Acknowledgement and DATA AVAILABILITY STATEMENTS}

This work was supported by the Gordon and Betty Moore Foundation EPiQS Initiative through Grant No. GBMF4534. The data that support the findings of this study are available from the corresponding author upon reasonable request.


\bibliography{ref}

\end{document}